\documentclass{article}

\usepackage{microtype}
\usepackage{graphicx}
\usepackage{subfigure}
\usepackage{booktabs} %

\usepackage[hyphenbreaks]{breakurl}
\PassOptionsToPackage{hyphens}{url}
\usepackage{hyperref}

\usepackage[accepted]{icml2024}

\usepackage{amsmath}
\usepackage{amssymb}
\usepackage{mathtools}
\usepackage{amsthm}

\usepackage[capitalize,noabbrev]{cleveref}

\theoremstyle{plain}

\theoremstyle{definition}

\theoremstyle{remark}

\usepackage[textsize=tiny]{todonotes}

\usepackage{fixmath}
\usepackage{paralist}
\usepackage[marginal,hang,stable]{footmisc}

\newcommand{\descr}[1]{\noindent\textbf{#1}}
\newcommand{\descrem}[1]{\noindent\emph{#1}}
\icmltitlerunning{Synthetic Data, SBPMs, and Regulatory (Non-)Compliance}

\begin{document}

\twocolumn[
\icmltitle{Synthetic Data, Similarity-based Privacy Metrics, and\\Regulatory (Non-)Compliance$^\dag$}

\icmlsetsymbol{equal}{*}

\begin{icmlauthorlist}
  \icmlauthor{Georgi Ganev}{ucl,hazy}
\end{icmlauthorlist}

\icmlaffiliation{ucl}{University College London, London, UK}
\icmlaffiliation{hazy}{Hazy, London, UK}

\icmlcorrespondingauthor{Georgi Ganev}{georgi.ganev.16@ucl.ac.uk}

\icmlkeywords{Machine Learning, ICML}

\vskip 0.3in
]

\renewcommand*{\thefootnote}{}
\footnotetext{$^\dag$For an extended version, refer to~\citep{ganev2023inadequacy}.\\}

\printAffiliationsAndNotice{}  %

\section{Motivation}

Synthetic tabular data, or data generated by machine learning generative models, is gaining popularity beyond academia and moving into real-world deployments.
Examples include releasing public census data by US~\citep{abowd2022the}, UK~\citep{ons2023synthesising}, and Israel~\citep{hod2024differentially}, as well as sharing sensitive financial and health data through private synthetic data vendors~\citep{ico2023synthetic, microsoft2024south}.
While these releases satisfy a formal definition of privacy, i.e., Differential Privacy (DP)~\citep{dwork2006calibrating}, this is still not the norm in numerous scientific papers~\citep{park2018data, lu2019empirical, zhao2021ctab, borisov2023language, yoon2023ehr, kotelnikov2023tabddpm, zhang2024mixedtype} and leading synthetic data vendors~\citep{mostly2020truly, syntegra2021fidelity, panfilo2022generating, syntho2023report}.
Instead, the papers/companies rely entirely on empirical ad-hoc privacy metrics based on the {\em similarity} between synthetic and real personal datasets.

\descr{Main Question.}
This prompts asking: {\em ``Is using similarity-based privacy metrics enough to consider synthetic data regulatory compliant?''}
Due to their fundamental issues and unreliable, inconsistent nature, {\em we argue that it is not}.

\vspace{-0.15cm}
\section{Definitions}

\begin{figure}[t!]
	\centering
	\subfigure{\includegraphics[width=0.95\linewidth]{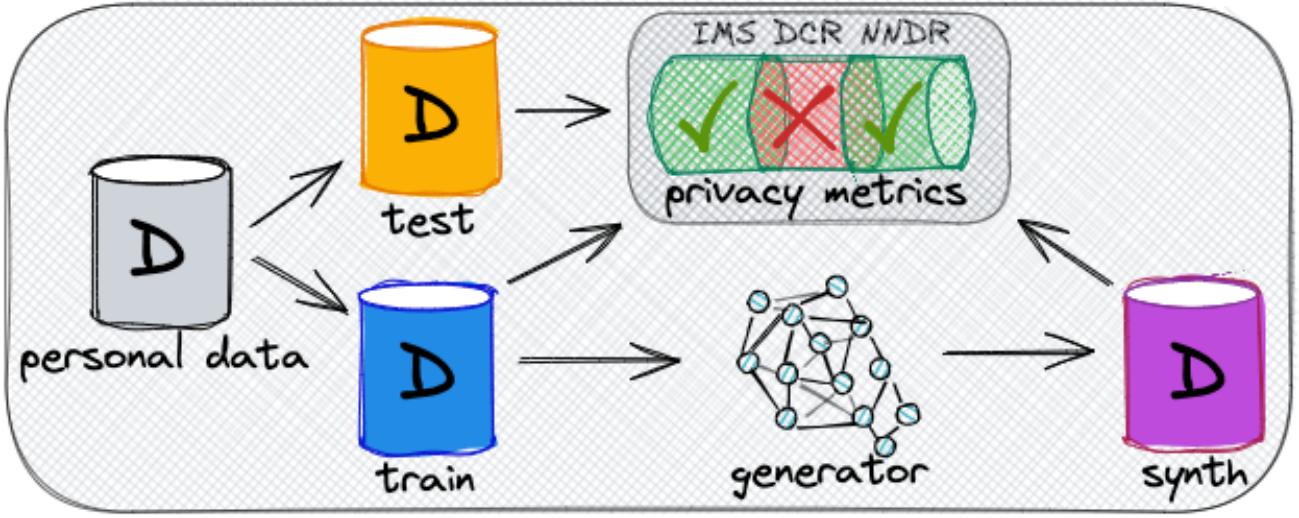}}
  \vspace{-0.1cm}
	\caption{Data flow overview.}
	\label{fig:reconsyn}
  \vspace{-0.4cm}
\end{figure}

\descr{Synthetic Data.}
We denote a real personal dataset as $\mathcal{D}$.
A generative model, $G$, is trained on $\mathcal{D}_{train}$ (a subset of $\mathcal{D}$; the remaining data, $\mathcal{D}_{test}$, is set aside for test purposes) to capture a probability representation, and could later be sampled to generate new (synthetic) data $\mathcal{D}_{synth}$ of arbitrary size (see bottom of Fig.~\ref{fig:reconsyn}).
Popular generative models include Graphical Models~\citep{zhang2017privbayes, mckenna2021winning}, GANs~\cite{xie2018differentially, jordon2019pate, xu2019modeling}, Diffusion Models~\cite{kotelnikov2023tabddpm, zhang2024mixedtype}, and Transformers~\cite{borisov2023language}.

\descr{Similarity-based Privacy Metrics (SBPMs).}
The intuition behind SBPMs is that $\mathcal{D}_{synth}$ should be representable and close to $\mathcal{D}_{train}$, but not closer than to $\mathcal{D}_{test}$~\citep{platzer2021holdout, mobey2022help}.
More precisely, the closest pairwise distances $d_{synth} = d(\mathcal{D}_{train}, \mathcal{D}_{synth})$ and $d_{test} = d(\mathcal{D}_{train}, \mathcal{D}_{test})$ are computed and their distributions compared through a statistical test (see Fig.~\ref{fig:reconsyn}).
The passing criterion is a comparison between a simple statistic run on each distribution, e.g., average/5th percentile~(p5).
In this paper, we focus on the three most widely used SBPMs by scientific papers and synthetic data vendors.
Finally, $\mathcal{D}_{synth}$ is deemed private if all three privacy tests pass~\citep{mostly2020truly, panfilo2022generating}.

\descrem{Identical Match Share (IMS)}
calculates the proportion of exact copies (statistic: average; test: $d_{synth} \leq d_{test}$).

\descrem{Distance to Closest Records (DCR)}
calculates the distances to their nearest neighbor in $\mathcal{D}_{train}$ (statistic: p5; test: $d_{synth} \geq d_{test}$).
DCR is meant to protect against scenarios where $\mathcal{D}_{train}$ is slightly perturbed and passed as $\mathcal{D}_{synth}$.

\descrem{Nearest Neighbor Distance Ratio (NNDR)}
follows DCR but divides the distances by the distance to their second neighbor (statistic: p5; test: $d_{synth} \geq d_{test}$).
The relative computations are supposed to further protect the outliers.

\begin{figure*}[t!]
	\centering
	\subfigure[$\mathcal{D}_{train}$ and $\mathcal{D}_{test}$]{\includegraphics[width=0.316\linewidth]{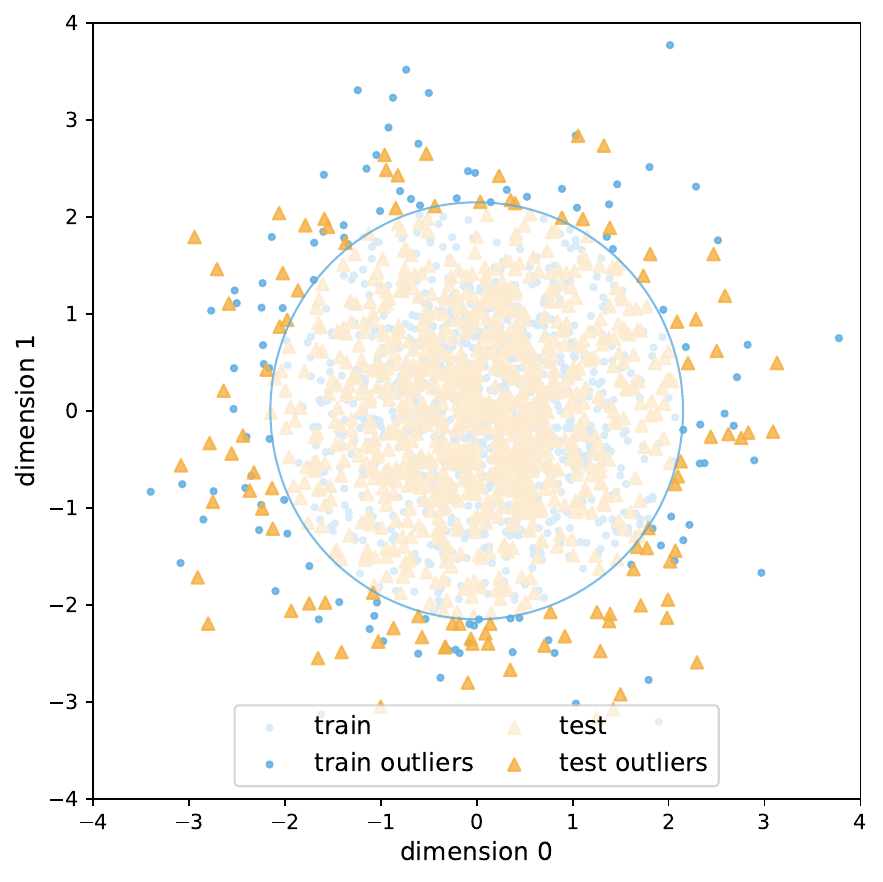}\vspace{-0.3cm}\label{fig:2d_gauss_train_test}}
	\subfigure[$\mathcal{D}_{synth}$ leaking all $\mathcal{D}_{test}$]{\includegraphics[width=0.31\linewidth]{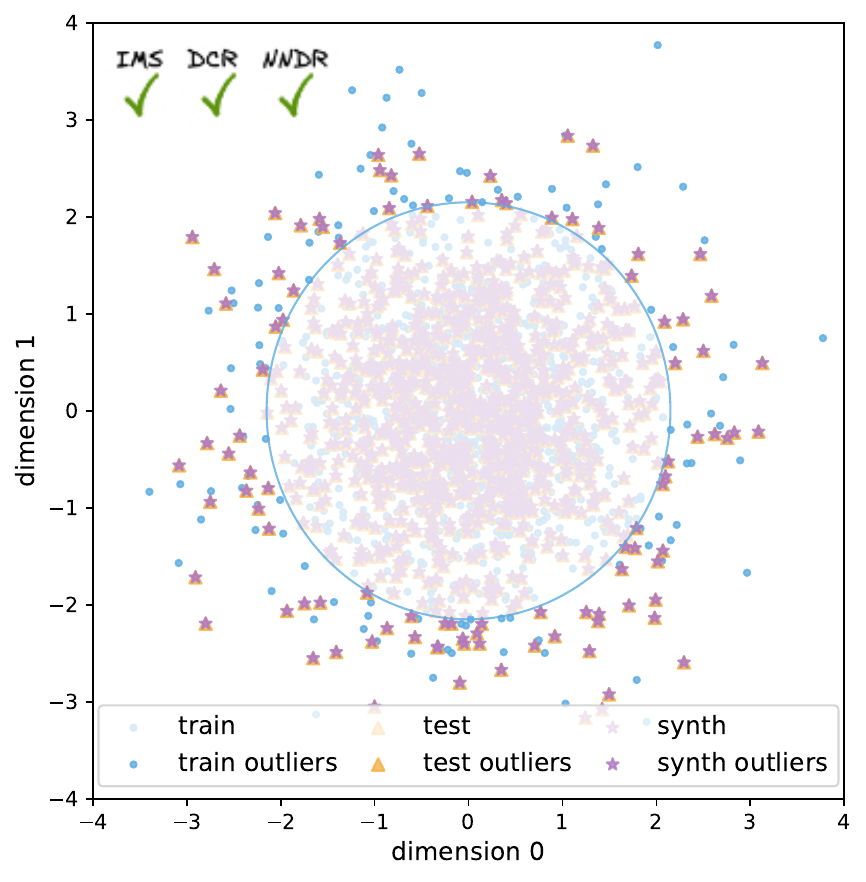}\vspace{-0.3cm}\label{fig:2d_gauss_test}}
	\subfigure[$\mathcal{D}_{synth}$ leaking $\mathcal{D}_{train}$ outliers]{\includegraphics[width=0.31\linewidth]{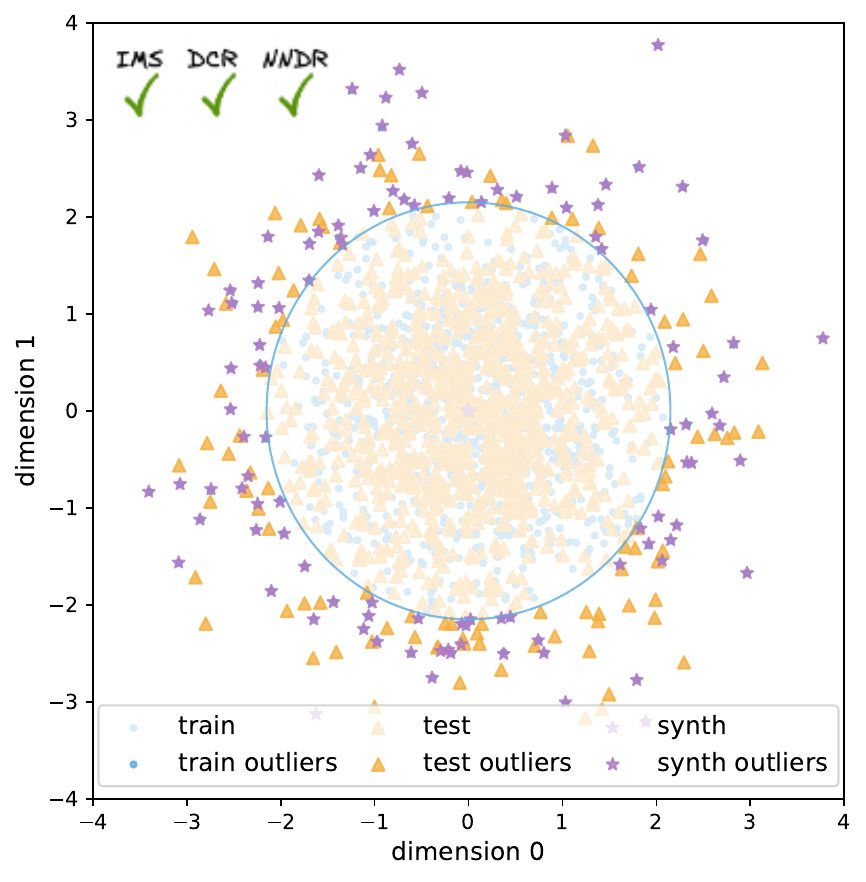}\vspace{-0.3cm}\label{fig:2d_gauss_train_out}}
  \vspace{-0.1cm}
  \caption{\emph{2d Gauss} data counter-examples.}
	\label{fig:2d_gauss}
  \vspace{-0.4cm}
\end{figure*}

\descr{GDPR.}
\citet{official2016article} define {\em personal data} as ``any information relating to an identified or identifiable living individual.''
Also, \citet{official2016recital} state that effectively anonymized information is not considered personal data and is exempt from data protection regulations.
Creating synthetic data from real personal data naturally involves processing it, so whether the result is personal or anonymous depends on the identifiability risk assessment.

\descrem{Sufficient Anonymization.}
``Effective anonymization reduces identifiability risk to a sufficiently remote level''~\citep{ico2021how}.
Assessing identifiability involves considering practical factors like cost, time, and technology, focusing on what is reasonably likely to be used rather than any theoretical possibility.
We mainly focus on two key technical risks for sufficient anonymization~\citep{eu2014opinion, ico2021how}: i) ({\em singling out}) isolating any individual, and ii) ({\em linkability}) combining records/datasets with synthetic data to identify an individual.
Last, \citet{ico2021how} proposes the {\em motivated intruder} test, suggesting that the risks should be assessed based on whether a competent intruder, with appropriate resources, could achieve identification if motivated enough.

\descr{Related Work.}
For more detailed discussion, refer to: synthetic data~\citep{jordon2022synthetic, de2024synthetic}, SBPMs~\citep{boudewijn2023privacy, ganev2023inadequacy, desfontaines2024empirical}, and regulation~\citep{lopez2022on, lopez2022synthetic, gal2023bridging, ganev2023when}.

\section{Fundamental Issues of SBPMs}

We identify several fundamental issues with using SBPMs to reason about privacy through empirical pass/fail tests.

\descr{No Theoretical Guarantees.}
SBPMs lack a defined threat model or strategic adversary, which ignores essential security~\citep{anderson2020security} and regulatory principles like the motivated intruder test.
Instead, they rely on arbitrarily chosen average-case statistics and held-out datasets, falling into the ``Generalization Implies Privacy'' fallacy~\citep{del2023bounding}, where generalization is average-case issue but privacy is a worst-case.
Thus, even if a model passes all tests and generalizes, it can still memorize data~\citep{song2017machine}.
Consequently, SBPMs offer no theoretical guarantees and are vulnerable to adversarial attacks.
Moreover, it is unclear whether or how the SBPMs correspond to the two technical risks -- singling out and linkability.

\descr{Privacy as Binary Property.}
SBPMs treat privacy leakage as binary, assuming one synthetic dataset is as safe as many (if the tests pass) even though $\mathcal{D}_{train}$ needs to be queried at each release.
However, the ``Fundamental Law of Information Reconstruction''~\citep{dwork2014algorithmic} warns that too many accurate answers can severely compromise privacy.

\descr{Privacy as Data Property.}
SBPMs see privacy as a property of the data, not of the generative model/process, which requires running the tests for each synthetic dataset.
This can lead to inconsistent results across generation runs.
Also, this increases privacy leakage as well as singling out and linkability risks.

\descr{Non-Contrastive Process.}
SBPMs do not compare computations with or without an individual.
Without noise or randomness, the system becomes vulnerable to attacks like differencing attacks, which, in turn, rules out plausible deniability and increases singling out concerns.

\descr{Incorrect Interpretation.}
The tests can be misinterpreted, as failing to reject the null hypothesis (``privacy is preserved'') does not confirm that privacy is indeed preserved.

\descr{Practical Issues.}
Most SBPMs implementations require discretizing the data, leading to imprecise calculations and overstated privacy protections. %
Also, the typical need for a 50/50\% train-test split can hurt the synthetic data quality. %

\section{SBPMs Counter-Examples}

We present three counter-examples showing the unreliability and inconsistency of SBPMs.
For all of them, we use a toy dataset, \emph{2d Gauss}, which consists of 2,000 points (split evenly between $\mathcal{D}_{train}$ and $\mathcal{D}_{test}$) drawn from a standard normal 2d distribution with no correlation (see Fig.~\ref{fig:2d_gauss_train_test}).
Approximately 10\% of the records, those outside the blue circle, are considered outliers.

\descr{Leaking All Test Data.}
Assume $\mathcal{D}_{synth}$ is an exact replica of $\mathcal{D}_{test}$ (Fig.~\ref{fig:2d_gauss_test}).
All privacy tests pass as $d_{synth}=d_{test}$.
Naturally, publishing half of the personal records cannot be considered regulatory compliant as this would leak an immense amount of private data and would fail the two technical risks -- singling out and linkability.

\descr{Leaking Train Data Outliers.}
Next, assume that $\mathcal{D}_{synth}$ contains all $\mathcal{D}_{train}$ outliers with small perturbations (the purple stars outside the circle in Fig.~\ref{fig:2d_gauss_train_out}) and numerous copies of the value (0, 0).
Again, all tests pass: there are no exact matches, and the (0, 0) values skew the distances enough to trick both DCR and NNDR.
Publishing $\mathcal{D}_{synth}$ will leave the outlier individuals unconvinced that their privacy is preserved~\citep{ons2018privacy, ico2022privacy} and would again fail the two technical risks.

\descr{SBPMs Inconsistency.}
We assume access to an oracle with knowledge of the generative process of $\mathcal{D}$.
Using the oracle, we sample 1,000 new $\mathcal{D}_{synth}$s.
Since no generative model was trained, i.e., $\mathcal{D}_{train}$ was never exposed, its privacy is preserved.
However, on only 274 occasions do all privacy tests pass, showing that the SBPMs fail to accurately capture the generating process.

The individual metric pass rates are 1 for IMS, 0.48 for DCR, and 0.38 for NNDR, revealing significant inconsistency.
Despite sampling from a fixed distribution, the metrics do not agree with each other while DCR and NNDR behave randomly.
Even if $G$ captures the underlying process without overfitting or memorization, the pass/fail outcome is sample-dependent, noisy, and unreliable.

Alternatively, if $\mathcal{D}_{synth}$ is fixed and $\mathcal{D}$ is randomly split into $\mathcal{D}_{train}$/$\mathcal{D}_{synth}$, we again run into inconsistencies.
Out of 1,000 such repetitions, only 380 cases pass all three tests.
This yet again demonstrates the inherent randomness and unreliability of the data-dependent train/test split and, in general, reasoning about privacy through SBPMs.

\section{Possible Countermeasures}

We discuss and disprove the efficacy of three intuitive solution intended to overcome the limitations of SBPMs.

\descr{DP Generative Models.}
The established framework to limit the ability of an attacker to exploit privacy leakage from trained models is to train them while satisfying DP.
Inspired by product deployments by synthetic data providers, we assume access to a single DP trained generative model and unperturbed metrics (per generation run).

In this scenario, although the likelihood of the model memorizing and reproducing real data records would be reduced, leakage can still occur when multiple $\mathcal{D}_{synth}$s are released along with the metrics.
The leakage comes from the privacy metrics themselves; since they require access to $\mathcal{D}_{train}$ and are deterministic (ruling out plausible deniability), they compromise the end-to-end DP pipeline.
Strategic and motivated adversaries could exploit this vulnerability to reconstruct real data records~\citep{ganev2023inadequacy}.
Adding additional privacy mechanisms on top of the metrics is unlikely to mitigate the problem, as the overall privacy integrity of the entire system needs to be carefully considered.

\descr{DP-fying the Metrics.}
Another possible solution could be to apply DP to the metrics.
However, this would not be robust.
Implementing DP to the metrics would require additional privacy budget for each generation run, which contradicts one of the main claimed advantages of adopting synthetic data: the ability to generate unlimited data.

\descr{Disabling Metrics Access.}
Finally, not disclosing the privacy metrics while still conducting statistical pass/fail tests would create significant issues.
First, users and customers would have to blindly trust the provider that the synthetic data meet a certain threshold.
Second, it would undermine a key selling point of providers: providing a tangible measure of compliance, essential for the product's transparency and explainability.
Additionally, with statistical pass/fail tests, sensitive information could still be vulnerable to present/future privacy attacks.

\section{Conclusion}

In this paper, we argue that SBPMs cannot ensure regulatory compliance of synthetic data.
SBPMs do not protect against singling out and linkability and, among other fundamental issues, completely ignore the motivated intruder test.

In~\citep{ganev2023inadequacy}, we discuss further fundamental issues of SBPMs, provide more SBPMs counter-examples, propose a novel reconstruction attack, which is capable of recovering the majority of $\mathcal{D}_{train}$ outliers, and argue that training DP generative models without access to SBPMs addresses their issues (note that DP comes with its own disadvantages, which we also discuss).

\descr{Empirical Evaluations.}
Privacy attacks and empirical evaluations should not be overlooked as they play a crucial role in identifying flaws, errors, and bugs in algorithms and implementations.
They contribute significantly to model auditing~\citep{jagielski2020auditing, nasr2023tight, annamalai2024what, ganev2024elusive} and improve the interpretability of theoretical privacy protections~\citep{houssiau2022framework, houssiau2022tapas}.

\section*{Acknowledgements}

We would like to thank Emiliano De Cristofaro for reviewing a draft of this paper and providing technical feedback.

%
%\bibliography{references}

\bibliographystyle{icml2024}

\end{document}